# Engineering and Tuning of high quality hexagonal boron nitride nanophotonic resonators

Otto Cranwell Schaeper[1,2], Angus Gale[1,*], Nathan Coste[1,2], Dominic Scognamiglio[1], Jonathan Horder[1], Hugo Quard[1,2], Igor Aharonovich[1,2,*]

[1] *School of Mathematical and Physical Sciences, Faculty of Science, University of Technology Sydney, Ultimo, New South Wales 2007, Australia*

[2] *ARC Centre of Excellence for Transformative Meta-Optical Systems (TMOS), University of Technology Sydney, Ultimo, New South Wales 2007, Australia*

* angus.gale@uts.edu.au, igor.aharonovich@uts.edu.au

*Abstract*

Van der Waals materials offer intriguing opportunities as building blocks for advanced quantum information technologies and integrated quantum photonic systems. Critical to their development are robust and high quality light-matter interactions, which can be delivered through the fabrication of optical resonators. Here we demonstrate a robust fabrication of one dimensional photonic crystal cavities (1D PCC) and microdisk resonators from hexagonal boron nitride, exhibiting Quality factors of ~ 4300 and ~ 8300, respectively. With these two resonator classes we demonstrated cavity mode tuning via atomic layer deposition and gas condensation. Cavity resonances were shifted by and ~9 nm for the 1D PCCs and ~16 nm in the microdisk resonators, respectively. Our work opens a promising pathway for a realisation of emitter – cavity coupling in hBN and eventually toward fully integrated quantum photonic circuitry with hBN.

Integrated photonics is a key enabler for quantum information processing [1, 2], sensing [3], and communication [4], enabling engineered light–matter interactions on-chip [5-7]. Across traditional three dimensional platforms—including for example silicon, diamond, lithium niobate, - mature fabrication technologies have delivered high-performance cavities and waveguides [8-12]. Performance of these devices is often assessed by measuring their quality factors (Q) and striving to engineer geometries with the smallest mode volume (V) [13-16].

Among explored cavity geometries, microdisk resonators support whispering-gallery modes (WGMs) with high Q and fields localized along the disk edge [17, 18]. These devices were often used to demonstrate strong coupling for ensembles of emitters and collective phenomena such as superradiance [18, 19]. While these microdisk resonators support multiple WGMs and exceptionally high Q factors, microdisk resonators have inherently large mode volumes. Alternative structures such as photonic crystal cavities (PCCs) can have extremely low mode volumes, thus making them attractive for quantum applications and coupling to single-photon sources [13, 20, 21]. The minimised mode volume is especially important in the context of emitter-cavity coupled systems, where Purcell Factor ($F_P$) ∝ Q/V, indicating the significant impact of V on the spontaneous emission rate. PCCs can have mode volumes <1 $(\lambda/n)^3$ allowing for extremely high theoretical $F_P$. Unlike microdisk resonators, PCCs typically support a single fundamental mode to maximize light–matter interaction with the embedded emitter. This, however, makes spectral alignment challenging due to fabrication variability of the PCCs and emitter wavelength dispersion. Consequently, post-fabrication tuning of cavity modes, for both microdisks and PCCs is therefore essential [22-24].

In recent years, hexagonal boron nitride (hBN) has emerged as a promising material for integrated photonics [25-30]. It is transparent in the visible to near infrared and hosts a variety of ultra bright single-photon emitters and spin-active defects, which can be engineered on demand with nanoscale resolution [31, 32]. Furthermore, hBN's van der Waals nature enables exfoliation into atomically smooth flakes of tuneable thickness, supporting fabrication of nanophotonic resonators and waveguides [33]. Due to the moderately low refractive index of hBN ($n \sim 1.8$) [34], optical isolation through undercutting is critical, as substrate leakage severely degrades cavity performance.

In the current work, we report the design, fabrication, and characterization of hBN-based PCCs and microdisk resonators, specifically designed for the blue and the visible spectral range. We are particularly focused on developing a robust undercutting process that so far was challenging to achieve with hBN. Our devices exhibit high Quality factors of ~ $Q \approx 4300$ at 578 nm for the PCCs, and $Q \approx 8300$ at 460 nm for the microdisk cavities, respectively. Further, we demonstrate cavity mode tuning for both types of resonators, with atomic layer deposition (ALD) modifying both PCC and microdisk resonances. Additionally, gas condensation was successfully used to demonstrate reversible tuning of the PCCs. ALD tuning is a global process, meaning that all cavities will be affected, but can exhibit increases in cavity Q factor [35, 36]. Gas tuning has limited scalability but presents a pathway toward integrated photonic devices through individual cavity addressability, with the ability to deterministically tune multiple components of a single device [16]. The renewed focus on single emitters in hBN, specifically the B-center [37] and spin-active single defects [38] show the necessity to both fabricate and tune high quality monolithic cavities, in order to enable emitter-cavity coupling. The record high Q factor from an hBN device thus far, and demonstration of multiple tuning methods, including the ability to shift WGM resonances across the full free spectral range (FSR), highlight the advances made toward hBN photonics in this work. Overall, these results demonstrate precise and reproducible spectral control of cavity modes in hBN devices—an essential step toward deterministic emitter–cavity coupling for scalable quantum photonic architectures.

Figure 1a shows a schematic illustration of the microdisk resonator and one dimensional (1D) PCC that are engineered and discussed in our work. The fabrication details are discussed in methods. Briefly, hBN was exfoliated onto silicon substrates, followed by patterning of the photonic structures using electron beam lithography (EBL) and a dry reactive ion etching. The fabricated structures were then suspended by undercutting the silicon in a potassium hydroxide (KOH) solution.

Figure 1(b-d) shows scanning electron microscope (SEM) images of the fabricated devices. Figure 1b displays two hBN microdisk resonators, sitting on a silicon pedestal. The surrounding hBN flake was inadvertently removed during the undercutting and drying process, but this has no effect on the efficacy of the device. Figures 1(c, d) display SEM images of 1D PCCs from different viewing angles, top down and tilted to 45 degrees, respectively. The 1D PCC features grating couplers to assist with the characterisation of the devices. The anisotropic nature of the silicon wet etch is clearly visible. Fabricated devices were aligned to ensure the etch would be deepest along the span of the cavity.

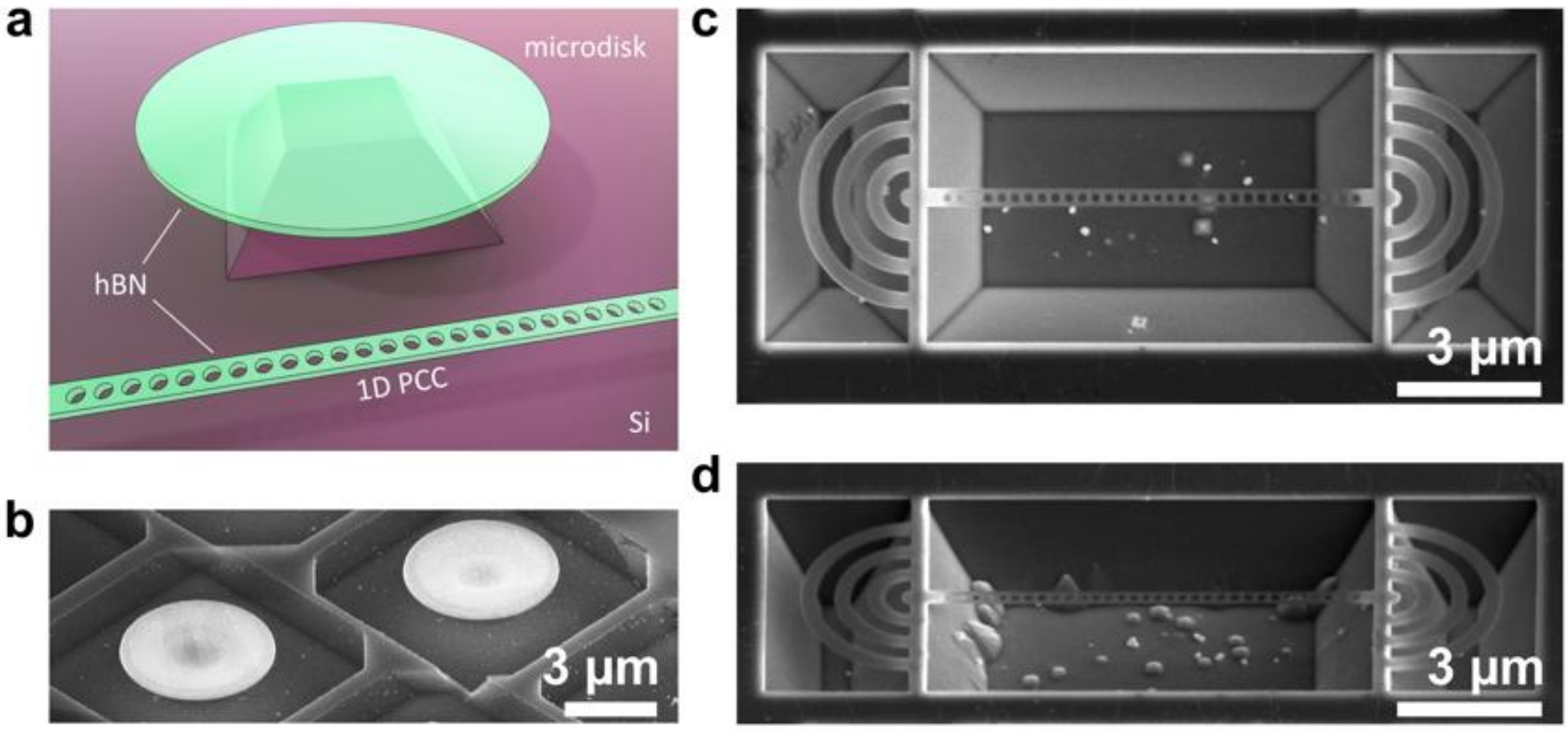

***Figure 1***. *Overview of the fabricated cavity designs. a) Schematic of the two monolithic cavity designs fabricated from hBN, namely a microdisk resonator and 1D photonic crystal cavity (PCC). The silicon substrate is etched to undercut the cavities. b) SEM image of 5 µm diameter hBN microdisks on silicon pedestals. c, d) SEM images of 1D PCCs integrated with grating couplers fabricated entirely from hBN. Top down and tilted to 45° angle, respectively.*

Figure 2(a,b) displays the Finite Difference Time Domain (FTDT) simulation of the cavity mode profiles simulated cavity. The mode confinement of 1D PCC is shown in two planes, (XY and XZ), respectively, with the cavity mode being confined to the central region of the device. The device characterisation was done using two complementary techniques. First, employing a standard photoluminescence (PL) measurement by confocal excitation and collection with a 532nm laser at the centre of the cavity. Second, by probing the cavity mode with resonant light: a broad laser pulse from a supercontinuum laser excites the cavity mode through a grating coupler, and the signal is collected from the top of the cavity. The excitation is polarized at a 45-degree angle relative to the cavity and the signal is collected in cross-polarization, in order to remove the light reflected on top of the structure (not coupling to the cavity). Both methods are shown schematically in Figure 2c. For both characterisation pathways the same device was measured, and spectra of the cavity mode were collected.

For the PL configuration, a $Q$ ~ 4272 was measured at ~ 585 nm, as shown in Figure 2e. Employing the cross polarization method, a slightly lower Q of 3071 at ~ 583 nm is observed, as shown in Figure 2d. The measured Q using the PL method is slightly higher since the excitation and collection are focused at the cavity center, so the mode is captured with minimal losses. In the cross-polarization configuration, excitation goes through the grating couplers, which introduce scattering and coupling losses, and the signal passes through polarization optics. If the cavity mode's polarization isn't perfectly aligned with the analyzer, this can distort the mode and broaden the resonance.

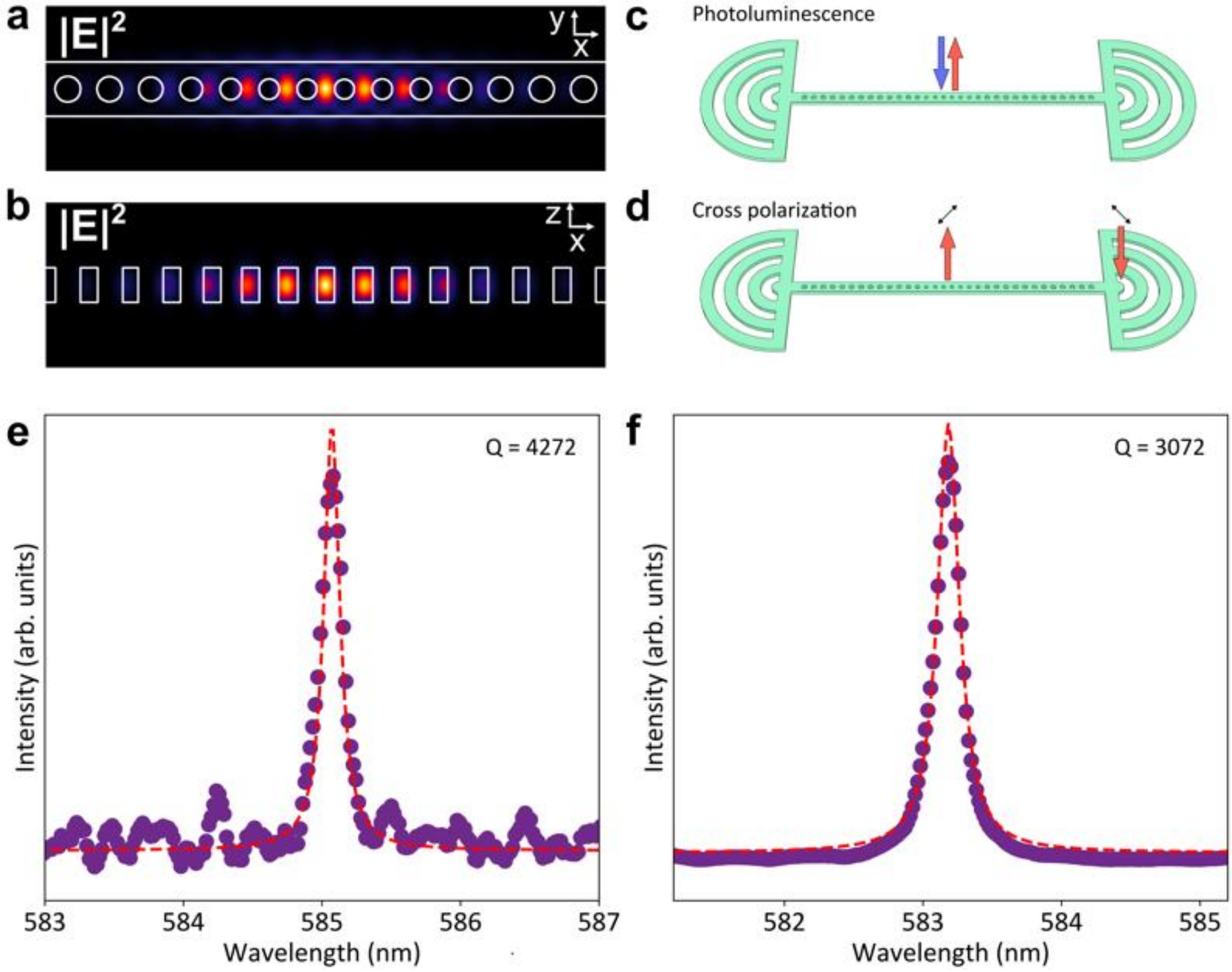

***Figure 2.*** *Characterisation of the 1D PCC resonant modes. a,b) FDTD simulations of the 1D PCC resonant mode at 577 nm in the XY (a) and XZ (b) planes. The mode is confined to the center of the cavity. c) Schematic showing the collection and excitation pathways to characterise the cavity modes employing standard photoluminescence (PL) and d) resonant cross polarization technique using grating couplers. The arrows indicate the polarization of the input and output pathways. e) PL spectrum of the highest Q mode measured from a 1D PCC, with a Q ~ 4272. f) Same cavity mode as in (e), measured using the cross polarization configuration, resulting in a Q of 3071.*

We now turn our attention to a controllable tuning of the cavity resonances. We implement two distinct approaches - namely gas condensation that modifies the cavity resonance due to change in refractive index, and ALD deposition that induces both a refractive index change and increases cavity physical thickness. We use ALD to deposit a few nm of aluminium oxide (AlO) on top of the cavity. After each run, the devices were measured using the cross polarized method as it results in a better signal to noise ratio.

Figure 3a shows the experimental results. The cavity resonance is red shifted after each deposition of $AlO_x$ ( 2 nm, 4 nm, 8 nm). As expected, a linear trend in red shifted signal is observed, with the increased thickness of $AlO_x$ as shown in figure 3b. Overall, we achieved a tuning range of ~ 16 nm without a significant degradation in the cavity quality factor (figure 3d, orange circles). Practically, high resolution cavity tuning of ~ 1 nm is possible employing this method by depositing sub nm $AlO_x$ films with ALD.

The second cavity resonance tuning approach attempted was employing a gas condensation. Here we loaded the 1D PCC into a cryostat, which was cooled down to 8K. This measurement was conducted using the cross polarization set up. We chose nitrogen as the condensation gas, and it was flown at a rate of 5 standard cubic centimeters per minute (sccm). During the flow, cavity resonances were recorded (each measurement took ~ 1 second). The cavity resonances then red shifted from ~714 nm to ~723 over ~100 seconds, as shown in Figure 3c. The mode profile broadened significantly as the mode red shifted, associated with a reduced quality factor.

The measured cavity quality factor, Q, as a function of tuning mechanism and duration are plotted in Figure 3d. For the ALD method, Q remained relatively stable across a 16 nm tuning range, with a maximum degradation of ~17%. On the other hand, upon the gas condensation, the Q factor degraded rapidly, by ~47% after 7 nm of resonance shift, most likely due to the ice condensation on the cavity and an increased absorption and distortion in the cavity's physical geometry.

The full dynamic tuning range of the cavity modes - both red shift (condensation) and blue shift (evaporation) are shown in Figure 3 (e, f). Initially the cavity mode shift is slow, most likely due to a delay between opening the mass flow controller and the gas entering the cryostat. The cavity mode at ~ 714 nm, then rapidly shifts towards a maximum wavelength of ~ 723 nm, representing a total shift of ~ 9 nm. The appeal of using gas condensation is the ease at which the cavity resonance can then be blue shifted back to its original resonance. This is achieved by sublimation of the ice through the introduction of a second laser. This laser is focused on the centre of the cavity, which warms it up and melts the condensate. Once all the ice has been removed the cavity mode is tuned back to the initial wavelength.

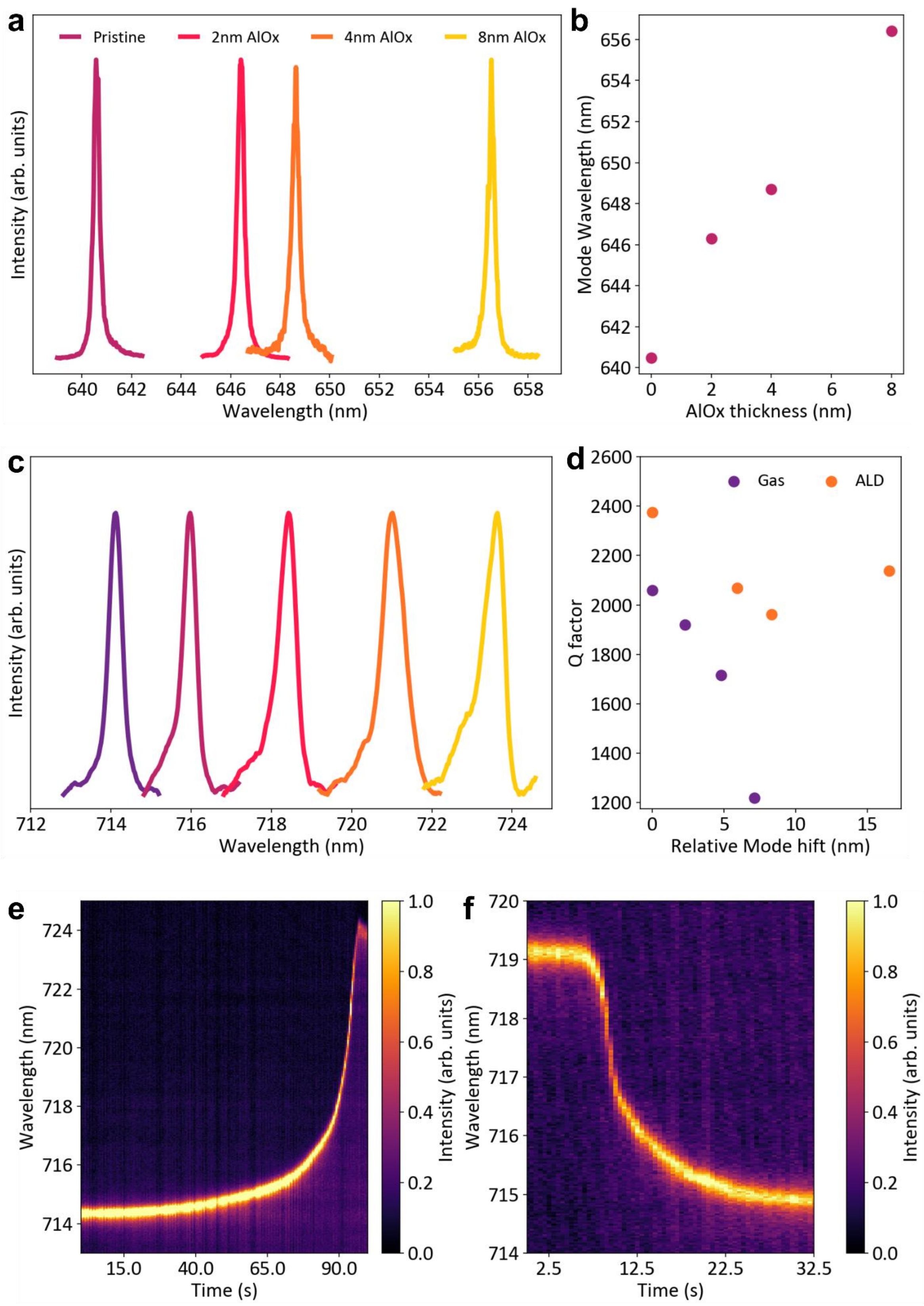

***Figure 3.*** *Cavity mode tuning methods for 1D PCCs. a) Cross polarization spectra from the same 1D PCC showing the resonant cavity mode before (pristine) and after ALD deposition of $AlO_x$ (2 nm, 4 nm, and 8 nm, respectively). b) Shift in the cavity resonance as a function of*

*$AlO_x$ thickness. c) Cross polarization spectra of a 1D PCC cavity mode tuned by a nitrogen gas condensation at 8K. The cavity mode is shown at different times as indicated in the figure. d) Measured cavity Q as a function of the spectral tuning from the initial cavity resonance for both ALD and gas tuning methods. e) Dynamic mode tuning of the 1D PCC in (c) by condensation, resulting in a red shift, and by sublimation (f), resulting in a blue spectral shift.*

We now turn our attention into tuning the WGMs of the microdisk resonators. High resolution SEM image of the microdisk, overlaid with FDTD modelling of the WGMs, is shown in figure 4a. The light region at the centre of the microdisk resonator is the pedestal, where the microdisk is connected to the silicon substrate below. To tune the cavity resonances of the microdisk, we implemented the ALD method. Figure 4b shows the WGMs recorded from this microdisk with a Q ~ 2800 at 635 nm and FSR of ~ 14 nm (purple curve). Upon deposition of $AlO_x$ the WGMs are red shifted. As we repeat the process with 2 nm, 4 nm, 8 nm and 20 nm of $AlO_x$, we achieve a shift across the full spectral range (as shown in Figure 4b). That means that the original WGM (indicated by the star), reaches the spectral location of the next WGM. Figure 4c displays the relationship between the amount of deposited $AlO_x$ to the WGM wavelength, exhibiting a linear trend.

Finally, we showcase the high *Q* observed at the blue spectral range, where the most promising quantum emitters in hBN are emitting [37, 39]. Figure 4d shows the WGMs recorded from a 5 μm diameter hBN microdisk, all exhibit Qs exceeding 6000 (SI Figure S6). The highest measured *Q* from this device was Q ~ 8286 at 460nm, as shown in figure 4e.

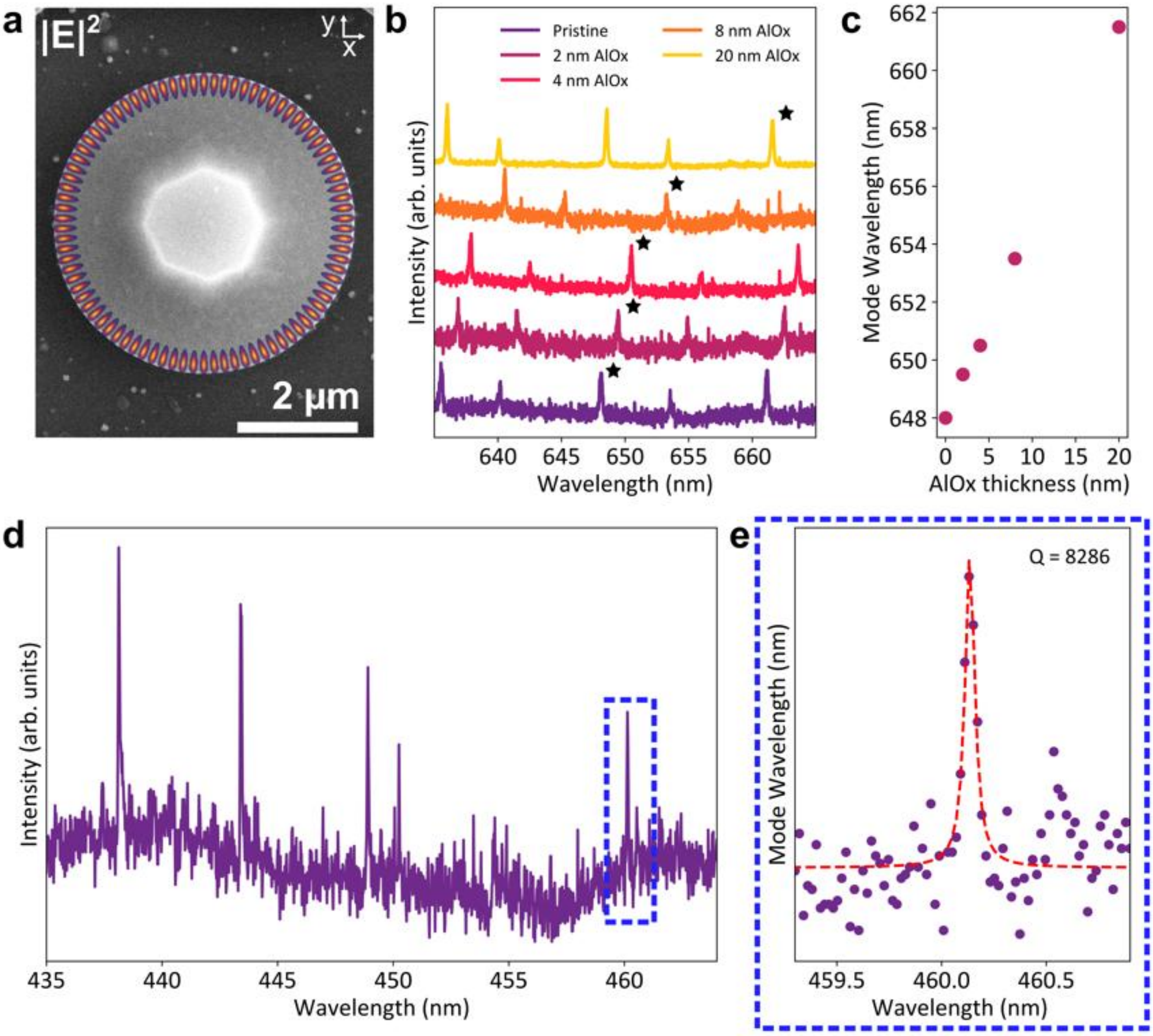

***Figure 4.*** *Microdisk cavity resonances and tuning. a) SEM image of the hBN microdisk cavity with a superimposed FDTD simulation of the mode profile at 457 nm. b) PL measurement of the WGMs from the same microdisk, showing the spectra before (pristine) and after ALD deposition of* $AlO_x$ *(2, 4, 8 and 20 nm). The spectra have been offset vertically for clarity. c) Shift in the cavity resonance as a function of* $AlO_x$ *thickness. d) WGMs of an hBN microdisk at the blue spectral range. e) Cavity mode highlighted in blue rectangle in (d) with a measured Q ~ 8282.*

A key aspect of the present work is the ability to reliably and controllably undercut the devices, thus enabling high Q cavities, specifically high Q WGMs in hBN microdisks. Earlier attempts of hBN device fabrication have almost exclusively used polymer hard masks for ICP-RIE processes - similarly to the method used in our work [40, 41]. However, recently, double etch methods employing a metal hard mask (chromium or nickel) have been demonstrated [42, 43]. These methods resulted in smoother side walls of the etched hBN nanostructures which is beneficial for optimal photonic component performance. In the development of this current work, we found that using a metal mask proved to be sub optimal for undercutting the underlying substrate. The undercutting was unreliable and produced irregular results with identical procedures, prompting a return to utilisation of a polymer mask for device fabrication, as undercutting was more reliable. The undercutting of 1D PCCs (or microdisk cavities) that were fabricated with a polymer mask takes approximately 12 minutes, when using 20% concentration KOH heated to 40°C. In comparison, we found that when using either a chromium or nickel double etch process, no clear undercut was achieved even after doubling or tripling the undercutting time. Raising temperature or bubbling the KOH is not ideal for hBN either, since the flakes can disassociate from the substrate. A selection of parameters used to undercut metal masked devices are stated in the SI in Table S1. Our results therefore suggest that if undercutting is required (i.e. for photonic cavities),a polymer mask is preferred. However, when devices can sit on a substrate (i.e.metasurfaces or waveguides), a metal mask can be used instead.

To conclude, we have successfully fabricated high quality photonic devices in hBN - specifically 1D PCCs and microdisk resonators. Through implementing a robust undercutting protocol, we achieved a record high Q from a monolithic hBN device in the blue spectral region, with qualities exceeding Q ~ 8,000. Further, we demonstrated two complementary cavity tuning mechanisms - gas condensation and ALD deposition - that can be readily implemented to tune cavity resonances in hBN. With the current focus on studying coherent properties of quantum emitters in blue spectral range, and the demonstrated here robust cavity engineering and tuning, we envision a promising pathway for a realisation of integrated quantum photonic circuitry with hBN.

## Acknowledgments

The authors acknowledge financial support from the Australian Research Council (CE200100010, FT220100053, DP250100973 ) the Air Force Office of Scientific Research (FA2386-25-1-4044). The authors thank the ANFF node at USYD and UTS for access to facilities.

The authors acknowledge Giorge Gemisis for his assistance, group politics keeping us apart.

## Methods

### Fabrication

hBN was exfoliated onto cleaned silicon substrates using scotch tape. Following this the samples were annealed in an STF1200 Tube Furnace. The tube furnace was programmed to ramp to 500C over 3 hrs where it remained for 4hrs, then to ramped to 650°C over 1 hr where it remained for 2 hrs then finally ramping to 820°C over 40 minutes where it remained for 40 minutes and then cooled down to ambient temperature over 6 hrs.

Following this, flakes were selected optically for fabrication. The samples were spin coated with an electron beam resist (CSAR.62 13) at 5K RPM for 60 seconds, and were then baked for 180 seconds at 180°C. The pattern was defined using electron beam lithography using an Elionix ELS-F125 for the 1D PCCs and a Helios G4 PFIB UXe DualBeam for the microdisk resonators . The pattern was developed by submerging the sample in CSAR developer 600-51, xylene and IPA for 60, 5 and 20 seconds respectively. A brief plasma cleaning step was conducted to remove residual polymer in the developed regions using a Trion inductively coupled plasma reactive ion etching (ICP-RIE) using the following parameters: 12 sccm oxygen, 10W ICP, 40W RIE, under 6 mTorr of pressure for 3 seconds.

Following this the pattern was transferred into the hBN in the same machine with the following parameters: 60 sccm Ar, 5 sccm SF6, 300W RIE under 10 mTorr of pressure. The residual CSAR was removed by submerging the sample in CSAR remover preheated to 135°C for one hour. Then rinsed in water and IPA before being blow dried using nitrogen.

For undercutting the silicon, the sample is submerged into 20% concentration potassium hydroxide preheated to 40°C for ~12 minutes to undercut both the 1D PCCs and microdisk resonators. Following this the devices were submerged in water and then IPA before being blow dried with nitrogen. Following this the device fabrication was completed.

**Photoluminescence**

Photoluminescent (PL) measurements were conducted on a home-built confocal microscope set up with a 532 nm continuous-wave laser excitation laser and a 100x objective (Nikon 0.9 NA). A 532 nm long-pass dichroic and 568 nm long-pass filter were used to eliminate any reflected laser into the collection. The collection and excitation position were overlapped in this configuration and positioned at the centre of the 1D PCCs, and on the edge of the microdisk resonators. Emission was collected using a spectrometer (Andor Kymera 328i) using the 1800 lines/mm grating.

Cross polarization measurements were conducted on a home-built optical set using a Fianium supercontinuum laser and a 100x objective (Nikon 0.9 NA). Emission was collected using a spectrometer (Andor Kymera 328i) using the 1800 lines/mm grating. In the collection excitation pathway, a $\lambda/4$ and $\lambda/2$ waveplates were used, rotated to maximise photon extinction on the spectrometer. The collection and excitation positions were uncoupled, with the collection at the centre of the cavity and the excitation on the grating couplers.